\documentclass[11pt,dvips]{article}

\usepackage{amsmath,amssymb,amsthm,amscd,latexsym}
\usepackage[toc,title,titletoc]{appendix}
\usepackage{cite}
\usepackage{graphicx,epsfig}
\usepackage[breaklinks,
colorlinks=true,
linkcolor=red,
urlcolor=blue]{hyperref}
\usepackage{breakurl}

\RequirePackage{color}

\allowdisplaybreaks[1]

\def\nnb{\\ \nonumber \\}
\def\nb{\nonumber &}
\def\nk{n_{\rm b}}

\def\rfr#1{eq. (\ref{#1})}

\def\dert#1#2{\frac{{{d}}{#1}}{{{d}}{#2}}}

\def\virg#1{``#1''}
\def\de{\right.}
\def\si{\left.}

\def\eqi{\begin{equation}}
\def\eqf{\end{equation}}
\def\eqia{\begin{eqnarray}}
\def\eqfa{\end{eqnarray}}
\def\Om{\mathit{\Omega}}
\def\rp#1#2{{#1\over#2}}
\def\lb#1{\label{#1}}

\def\Kx{\hat{K}_x}
\def\Ky{\hat{K}_y}
\def\Kz{\hat{K}_z}
\def\bds#1{\boldsymbol{#1}}

\def\co{\cos\omega}
\def\so{\sin\omega}

\def\cO{\cos\Om}
\def\sO{\sin\Om}

\def\cI{\cos I}
\def\sI{\sin I}

\def\ee{e^2}

\def\ton#1{\left(#1\right)}
\def\qua#1{\left[#1\right]}
\def\grf#1{\left\{#1\right\}}

\begin{document}

\title{Modified theories of gravity with nonminimal coupling and \textcolor{black}{orbital particle} dynamics}

\author{L. Iorio\\ Ministero dell'Istruzione, dell'Universit$\grave{\textrm{a}}$ e della Ricerca (M.I.U.R.)-Istruzione \\ Fellow of the Royal Astronomical Society (F.R.A.S.)\\ Viale Unit$\grave{\textrm{a}}$ di Italia 68, 70125, Bari (BA), Italy}

\maketitle

\begin{abstract}
 We consider a  non-rotating, massive test particle acted upon by a \virg{pressure}-type, non-geodesic acceleration arising from a certain general class of gravitational theories with nonminimal coupling between the matter and the metric. The resulting orbital perturbations for a two-body  system are investigated both analytically and numerically. \textcolor{black}{Remarkably}, a secular increase of the two-body relative distance occurs. In principle, it may yield a physical mechanism for the steady recession of the Earth from the Sun recently proposed to explain the Faint Young Sun Paradox in the Archean eon.
 At present, the theorists have not yet derived explicit expressions  for some of the key parameters of the model, such as the integrated \virg{charge} $\xi$, depending on  the matter distribution of the system, and the 4-vector $K^{\mu}=\{K^0,\bds K\}$ connected with the nonminimal function $F$. Thus, we phenomenologically treat them as free parameters, and preliminarily infer some indications on their admissible values according to the most recent Solar System's planetary ephemerides. From the latest determinations of the corrections $\Delta\dot\varpi$ to the standard perihelion precessions,  estimated by the astronomers who produced the EPM2011 ephemerides without modeling the theory considered here, we preliminarily obtain $|\xi K|\lesssim 0.1$ kg s$^{-1}$ for \textcolor{black}{the Sun and} Mars. From guesses on what could be the current  bounds on the secular rates of change of the planetary semimajor axes, we get $|\xi K^0|\lesssim 1249$ kg s$^{-1}$ for Mars. More effective constraints could be posed by reprocessing the same planetary data sets with dedicated dynamical models including the effects studied here, and explicitly  estimating the associated parameters. The \textcolor{black}{Earth and} the COBE and GP-B satellites yield $|\xi K|\lesssim 2\times 10^{-4}$ kg s$^{-1}$ and $|\xi K\textcolor{black}{^0}|\lesssim 2\times 10^{-10}$ kg s$^{-1}$, respectively.
\end{abstract}



\centerline
{PACS: 04.50.Kd; 95.10.Ce; 04.80.-y; 95.10.Km}

\section{Introduction}\lb{Introduzione}
In this paper, we investigate the orbital effects of the covariant equations of motion for \textcolor{black}{structureless massive} test bodies in gravitational theories with general nonminimal coupling between the matter and the spacetime metric \cite{2013PhRvD..87d4045P,2013arXiv1307.3933P} by considering a local, gravitationally bound two-body system. In particular, we will consider the case in which the nonminimal coupling is allowed to be  a general function of the set of 9 parity-even curvature invariants \cite{2013PhRvD..87d4045P,2013arXiv1307.3933P}. \textcolor{black}{For other generalized theories with nonminimal coupling functions depending on curvature invariants, see, e.g., \cite{2007PhRvD..75j4016B,2008PhRvD..78f4036B,2011PhR...505...59N} and references therein}.

\textcolor{black}{The authors of \cite{2013PhRvD..87d4045P,2013arXiv1307.3933P} derive the equations of motion for (extended) test bodies from the energy-momentum conservation law by adopting the covariant multipolar approximation scheme \cite{Dixon64} utilizing the expansion technique by Synge \cite{Synge60}. Although such an expansion can be carried out to any multipolar order \cite{2013PhRvD..87d4045P}, here we confine to the lowest order case.
As such,} the  non-geodesic four-acceleration of a  \textcolor{black}{non-rotating, structureless  massive} test particle is\footnote{In \cite{2013PhRvD..87d4045P,2013arXiv1307.3933P}, the Latin letters are used for the spacetime indices.} \cite{2013PhRvD..87d4045P,2013arXiv1307.3933P}
\eqi A^{\alpha} = \rp{c\xi}{\mathfrak{m}}\ton{\delta^{\alpha}_{\beta} -\rp{v^{\alpha}v_{\beta}}{c^2}}K^{\beta},\ \alpha=0,1,2,3.\lb{4accel}\eqf
In \rfr{4accel}, \textcolor{black}{which corresponds to eq. (55) of \cite{2013PhRvD..87d4045P} and to eq. (11) of \cite{2013arXiv1307.3933P}}, $c$ is the speed of light in vacuum, \textcolor{black}{$\mathfrak{m}$ is the mass of the test particle as defined in  multipolar schemes within the general relativistic framework},  $\delta_{\beta}^{\alpha}$ is the four-dimensional Kronecker delta, $v^{\alpha}$  is the four-velocity of the test particle, \textcolor{black}{the \virg{charge} $\xi$} is an integrated quantity depending on the matter distribution of the system, $K^{\alpha}\doteq \nabla^{\alpha}\ln F,$ where $\nabla^{\alpha}$ denotes the covariant derivative, and the nonminimal function $F$ depends arbitrarily on the spacetime metric $g_{\alpha\beta}$ and on the Riemann curvature tensor $R_{\alpha\beta\gamma}^{\ \ \ \ \delta}$. It must be noticed that \rfr{4accel} is given in covariant form \cite{2013PhRvD..87d4045P}; no assumptions have been made regarding the coordinates. Taking \rfr{4accel} as a starting point, one can make use of any coordinate system one likes.
Concerning $\xi$, its general definition is \cite{2013PhRvD..87d4045P}
\eqi\xi\doteq\int_{\Sigma(s)}{\mathcal{L}_{\rm mat}}w^{x_2}d\Sigma_{x_2},\eqf where $\mathcal{L}_{\rm mat}$ is the matter Lagrangian density, $s$ is the particle's proper time, and the integral is performed over a spatial hypersuperface; for details on the index notation adopted and on the explicit calculation of $w^{\alpha}$, see  \cite{2013PhRvD..87d4045P} and the appendix of \cite{Dixon64}, respectively. In general, $\xi$ would contain contributions from both a background source and from the test particle. To this aim, see in particular the general form of the conservation law of eq. (34) in  \cite{2013PhRvD..87d4045P}. However, in a multipolar context, in particular in \rfr{4accel}, one would consider only test particles in source-free regions; thus, $\xi$ would correspond to the test particle only.

Knowing the orbital effects of \rfr{4accel} on the dynamics of massive bodies is particularly important to put  theories with nonminimal coupling on the test. Indeed, as recently pointed out in \cite{2013arXiv1307.3933P}, they cannot be probed by massless photons since they move along null geodesics.
\section{The orbital effects on the motion of a non-rotating test particle}\lb{orbite}
From \rfr{4accel}, the test particle acceleration\textcolor{black}{\footnote{\textcolor{black}{Here and in the following, \virg{nmc} stands for \virg{non-minimal coupling}.}}}
\eqi{\bds A}_{\rm nmc} =  -\rp{\xi\qua{c^2 \bds K - c K\textcolor{black}{^0} \bds v +\ton{\bds K\bds\cdot\bds v}\bds v }}{c \mathfrak{m}}\lb{accel},\eqf written in the usual three-vector notation, can be obtained. \textcolor{black}{It should be thought as an additional, body-dependent acceleration with respect to  the Newtonian monopole and the general relativistic acceleration including, e.g., the 1PN Schwarzschild and Lense-Thirring terms. In view of the excellent agreement among observations and their standard general relativistic models in a variety of astronomical and astrophysical systems \cite{2001LRR.....4....4W}, it is reasonable to look at \rfr{accel} as a small perturbation of the usual general relativistic geodesic acceleration.}

As such, the impact of \rfr{accel} on the orbit of a test particle can be analytically worked out with the standard Gauss equations \cite{Bertotti03} which allow to treat any kind of perturbative acceleration, independently of its physical origin. Since we are interested in the long-term effects, we need to integrate the r.h.s. of the Gauss equations, evaluated onto the Keplerian ellipse chosen as unperturbed reference orbit, over one orbital revolution.
In doing that, we only consider, to first order, the largest contributions coming just from the interplay between the Newtonian acceleration and the non-geodesic one.
%
In the calculation, as a working hypothesis, we make the assumption that $\mathfrak{m},\xi,K^{\alpha}$ can be considered as uniform over the spatial extension $L$ of the two-body system  considered and constant during its characteristic timescale set by the orbital period $P_{\rm b}$. \textcolor{black}{To this aim, it is important to stress that, in general, $\mathfrak{m},\xi,K^{\alpha}$ are not constant  for the very general class of theories covered in \cite{2013PhRvD..87d4045P}; thus, in principle, assuming  their constancy, even over one orbital period of the test particle, would need further justification.}
\subsection{Analytical calculation}\lb{calcolo}
We will not make any further a-priori assumptions on either the orbital configuration of the test particle or the spatial orientation of $\bds K$, determined by the components  $\Kx,\Ky,\Kz$ of the unit vector $\bds K/K$.

A lengthy calculation allows to obtain the averaged rates of change of\footnote{While $a$ and $e$ characterize the size and the shape ($0\leq e <1$) of the unchanging Keplerian ellipse, respectively, $I,\Om,\omega$ can be thought as three Euler angles determining its orientation in the inertial space.} the semimajor axis $a$, the eccentricity $e$, the inclination $I$, the longitude of the ascending node $\Om$ and the argument of pericenter $\omega$. In the following, $\nk\doteq\sqrt{GM a^{-3}}=2\pi/P_{\rm b}$ is the unperturbed Keplerian mean motion, where $G$ is the Newtonian constant of gravitation and $M$ is the mass of the source body considered as the primary in the two-body system considered.

The velocity-independent term in \rfr{accel} yields
\begin{align}
\dert a t \lb{dadt1} & = 0,\nnb
\dert e t \lb{dedt1} \nb = -\rp{c\xi K}{32 \mathfrak{m} a\nk}\mathcal{E}\ton{e}\qua{\Kz\sI\co + \cI\co\ton{\Ky\cO - \Kx\sO} - \de\nnb
&\si - \so\ton{\Kx\cO + \Ky\sO} },\nnb
\dert I t \lb{dIdt1} & = \rp{3c\xi K e \co }{2 \mathfrak{m} a  \nk\sqrt{1 - \ee}}\qua{\Kz\cI +\sI\ton{\Kx\sO - \Ky\cO}},\nnb
\dert\Om t \lb{dOdt1} & =
\rp{3c\xi K e \csc I\so }{2 \mathfrak{m} a  \nk\sqrt{1 - \ee}}\qua{\Kz\cI +\sI\ton{\Kx\sO - \Ky\cO}}, \nnb
\dert\omega t  \lb{dodt1} \nb = -\rp{c\xi K }{8 \mathfrak{m} a \nk }\mathcal{P}\ton{e}\grf{\co\ton{\Kx\cO + \Ky\sO} + \de\nnb
\nb\si +\so\qua{\Kz\sI +\cI\ton{\Ky\cO - \Kx\sO}  }  }-\nnb
& -  \cI\dert\Om t,
\end{align}
with
\begin{align}
\mathcal{E}\ton{e} \nb = -\rp{\sqrt{1 - \ee}}{\ee}\grf{116 e^6 + 48 e^7 + 12 e^8 -16e\ton{1-e}\sqrt{1-\ee} +\de\nnb
\nb\si + e^5\qua{104 + 3\ton{1-e}\sqrt{1-\ee}} + \de\nnb
\nb\si + e^3\qua{32+5\ton{1-e}\sqrt{1-\ee}}+\de\nnb
\nb\si + e^2\qua{-32 + 5\ton{1-e}\sqrt{1-\ee} - 5\ton{1-\ee}^{3/2} } + \de\nnb
\nb\si + e^4\qua{40 + 3\ton{1-e}\sqrt{1-\ee} - 3\ton{1-\ee}^{3/2} } +\de\nnb
\nb\si + 16\qua{-\ton{1-e}\sqrt{1-\ee} + \ton{1-\ee}^{3/2} }}\approx \nnb
& \approx 32\left(1-e\right) - 56\ee + \mathcal{O}\ton{e^3},\nnb
\mathcal{P}\ton{e} \nb = \textcolor{black}{-\rp{1}{e}\sqrt{\rp{1+e}{1-e}}\ton{15 - 7e + 5 \ee + 3 e^3 -12 e^4 -4 e^5}} \approx\nnb
&\approx -\rp{15}{e} -8 -\rp{11e}{2} -12\ee + \mathcal{O}\ton{e^3}.
\end{align}

The term  \textcolor{black}{in \rfr{accel} that is linear in $v$} affects only the semimajor axis $a$ according to
\begin{align}
\dert a t \lb{dadt2} & = \rp{2\xi K\textcolor{black}{^0}}{\mathfrak{m}}a, \nnb
\dert e t \lb{dedt2} & = 0, \nnb
\dert I t \lb{dIdt2} & = 0, \nnb
\dert\Omega t \lb{dOdt2} & = 0, \nnb
\dert\omega t \lb{dodt2} & = 0.
\end{align}

The long-term effects due to the term of order $\mathcal{O}\ton{v^2}$ in \rfr{accel} are
\begin{align}
\dert a t \lb{dadt3} \nb = \rp{4 \xi K \nk a^2}{c \mathfrak{m}}\mathcal{A}\ton{e}\qua{\Kz\sI\co + \cI\co\ton{\Ky\cO - \Kx\sO} - \de\nnb
&\si - \so\ton{\Kx\cO + \Ky\sO} },\nnb
\dert e t \lb{dedt3} \nb = \rp{2 \xi K \nk a}{c \mathfrak{m}}\mathcal{E}^{'}\ton{e}\qua{\Kz\sI\co + \cI\co\ton{\Ky\cO - \Kx\sO} - \de\nnb
&\si - \so\ton{\Kx\cO + \Ky\sO} },\nnb
\dert I t \lb{dIdt3} & = 0,\nnb
\dert\Om t \lb{dOdt3} & = 0, \nnb
\dert{\textcolor{black}{\omega}} t \lb{dodt3} \nb =\rp{2\xi K \nk a}{c \mathfrak{m} }\mathcal{P}^{'}
\ton{e}\grf{\co\ton{\Kx\cO + \Ky\sO} + \de\nnb
&\si +\so\qua{\Kz\sI +\cI\ton{\Ky\cO - \Kx\sO}  }  },
\end{align}
with
\begin{align}
{\mathcal{A}}\ton{e} \lb{Ae} &= \rp{1 - \ee - \sqrt{1 - \ee}}{e\sqrt{1 - \ee}}\approx -\rp{e}{2} + \mathcal{O}\ton{e^3},\nnb
{\mathcal{E}}^{'}\ton{e} \lb{Ee} & = \rp{-1 + \ee}{1 + \sqrt{1 - \ee}}\approx -\rp{1}{2} +\rp{3\ee}{8} +\mathcal{O}\ton{e^4},\nnb
{\mathcal{P}}^{'}\ton{e} \lb{Fe} & =\rp{-1 + \ee +  \sqrt{1 - \ee}}{e^3}\approx \rp{1}{2e} - \rp{e}{8} + \mathcal{O}\ton{e^3}.
\end{align}
An important feature of \rfr{accel} is that it causes long-term changes of both the semimajor axis $a$ and the eccentricity $e$, thus impacting the two-body mean distance $\overline{r} = a\ton{1 + \ee/2}$ as well. \textcolor{black}{It should be noted that \rfr{dadt2} and \rfr{dadt3} do not allow   to establish a-priori  the repulsive or attractive character of the modification of the orbit's size. Indeed, if on the one hand it depends on the orbital configuration of the particle through $e,I,\omega,\Om$ (\rfr{dadt3}, \rfr{Ae}), on the other hand also the signs of $K,\xi,\mathfrak{m}$ are, in principle, relevant. At the present stage of the development of the class of theories with non-minimal coupling considered here  \cite{2013PhRvD..87d4045P,2013arXiv1307.3933P},  the analytical form of $K^{\alpha},\xi,\mathfrak{m}$ for a localized two-body system have not yet been explicitly worked out. It will certainly be a necessary and important step forward. It is worthwhile remarking also that there are no Newtonian perturbing accelerations of a detached  binary able to secularly impact on $a$. Concerning standard general relativity, actually there are two effects that, at the 1PN level, can in principle change both $a$ and $e$: the temporal variation of the masses entering the gravitational parameter $\mu\doteq GM$ of the two-body system \cite{1991ercm.book.....B,2010SRXPh2010.1249I}, and the cosmological expansion \cite{2012PhRvD..86f4004K,2013MNRAS.429..915I}. Nonetheless, being proportional to $\dot\mu$ \cite{2010SRXPh2010.1249I} for the specific system considered and to the Hubble parameter $H$ \cite{2012PhRvD..86f4004K,2013MNRAS.429..915I}, they do not contain any free parameter: their magnitude is completely negligible in any realistic astronomical scenario of interest.  }
\subsection{Numerical calculation}\lb{numero}
We confirmed the results obtained analytically in Section \ref{calcolo} by performing a numerical integration of the equations of motion, in cartesian coordinates, of an Earth-like test particle orbiting a Sun-sized source body initially at 1 au  under the influence of the Newtonian monopole and of the non-geodesic acceleration of \rfr{accel}. The time span of the integration was chosen much longer than the unperturbed orbital period of the test particle. For the sake of simplicity, we adopted  $e=I=0$ as initial values for the eccentricity and the inclination of the test particle, along with $\Kx=\Ky=0$.  According to \rfr{dadt1}-\rfr{dodt3}, the only non-vanishing secular change occurs for the semimajor axis, as per \rfr{dadt2}. Thus, the distance $r$ from the primary should experience a steady, cumulative increase. Figure \ref{figura1}, displaying the difference $\Delta r$ between the numerically integrated distances with and without \rfr{accel} for the same initial conditions, confirms such a prediction. In order to make a meaningful confrontation with the perturbative calculation of Section \ref{calcolo},  we choose the values of $\xi, K^0, K$ in our numerical integration in such a way that the resulting non-geodesic acceleration was several orders of magnitude smaller than the Newtonian one. The units used in Figure \ref{figura1} and the magnitude of the effect depicted in it are not to be considered indicative.
\begin{figure*}
\centering
\begin{tabular}{c}
\epsfig{file=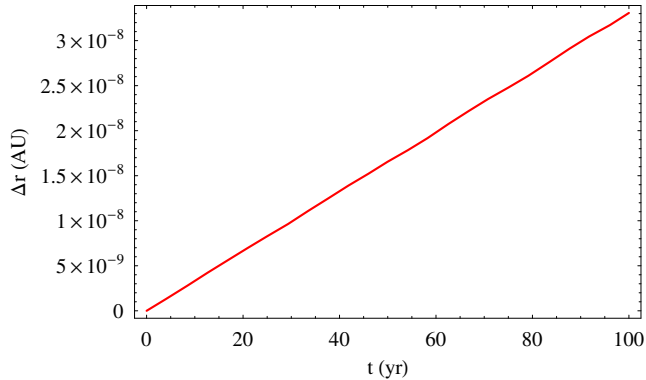,width=0.70\linewidth,clip=}
\end{tabular}
\caption{Difference $\Delta r$ between the numerically integrated time series of the distance with and without \rfr{accel}. \textcolor{black}{The equations of motion were integrated with a numerical solver using explicit Runge-Kutta methods.} We adopted a Sun-like star as primary and an Earth-sized planet as test particle. As far as the parameters $\xi$ and $K^{\alpha}$ entering \rfr{accel} are concerned, we adopted $\Kx = \Ky =0$, while their magnitudes were chosen in order to have an initial ratio $A_{\rm nmc}/A_{\rm Newton}\approx  10^{-7}$ for \rfr{accel} and the Newtonian acceleration; as far as their sign is concerned, the positive one was taken. As initial conditions of the test particle, common to both the numerical integrations, we adopted $x_0 = a(1-e), y_0=z_0=0, \dot x_0 = 0, \dot y_0 = \sqrt{GM a^{-1}(1+e)(1-e)^{-1}},\dot z_0 =0$, with $a=1$ au, $e=I=0$. The time span is $\Delta t = 100$ yr. \textcolor{black}{The resulting secular increase is in agreement with the theoretical prediction of \rfr{dadt2} when calculated with the same physical and orbital parameters adopted in the numerical integration.}
}\lb{figura1}
\end{figure*}
Finally, we mention the fact that the variation of the distance of the two-body system implied by the rates of change of $a$ and $e$ may be used, at least in principle, to accommodate  the required orbital recession of the Earth away from the Sun postulated in \cite{2013Galax...1..192I} to explain the so called Faint Young Sun Paradox \cite{2012RvGeo..50.2006F} during the Archean eon.
\section{Preliminary confrontation with the observations}
From the point of view of a possible confrontation with the observations,  we generally remark that, at this stage of the development of the theory considered in this paper, no explicit expressions for the integrated charge $\xi$ have been worked out for a localized two-body system. Similar considerations also hold for $K^{\alpha}$, which should be fixed by making some specific choices for the nonminimal function $F$. Thus, \textcolor{black}{strictly speaking,} it is not yet possible to quantitatively predict the magnitude of the orbital effects computed in Section \ref{calcolo} for, say, a Sun-planet \textcolor{black}{or Earth-satellite} pairs.
Nonetheless,  the inverse approach can be adopted, in principle, by treating $\xi$ and $K^{\alpha}$ as free parameters to be phenomenologically constrained or determined from existing data records processed for different purposes.
\textcolor{black}{
To this aim, some considerations are in order\footnote{I am grateful to an anonymous referee for the following considerations.}.  In \cite{2013PhRvD..87d4045P}, the metric is sourced by matter, which it couples to via the right hand side of eq. (9) in \cite{2013PhRvD..87d4045P}. Different terms in these non-minimal couplings appear to contain different number of derivatives on the metric. This fact may likely imply that terms with fewest number of derivatives should be dominant in the post-Newtonian limit. It could be  expected that these dominant\footnote{The dominant terms should also be examined to see exactly what order in the PN expansion they would become relevant.} terms  introduce new mass scales in the problem other than $G \sim 1/M_{\rm P}^2$. It is just these scales that ought to be constrained when compared to observations.
}
\textcolor{black}{\section{Bounds from Solar System's planetary motions}\lb{soles}}
The recently determined supplementary precessions $\Delta\dot\varpi$ of the perihelia\footnote{It turns out that the supplementary precessions of the nodes $\Delta\dot\Om$ \cite{2011CeMDA.111..363F} are less effective for our purposes.}  of some planets of our Solar System \cite{Pitjevi013,2013MNRAS.432.3431P} are used here. Since such corrections $\Delta\dot\varpi$ to the standard Newtonian-Einsteinian perihelion precessions, which globally account for all the dynamical and measurement modeling errors, are statistically compatible with zero, they can be suitably used to place some \textcolor{black}{bounds} on $|\xi K|$  by comparing them with the analytical precessions of Section \ref{calcolo}.
\textcolor{black}{T}he non-standard dynamics treated in this paper was \textcolor{black}{not} explicitly modeled in \cite{2011CeMDA.111..363F,Pitjevi013,2013MNRAS.432.3431P}, \textcolor{black}{so that no} dedicated parameters such as $\xi,K^{\nu}$ were estimated \textcolor{black}{in a least-square way} in dedicated covariance analyses along with, say, other parameters taking into account further potential deviations from general relativity. Nonetheless, the bounds which will be inferred here
are useful since they are an indication of acceptable values offered by the most recent planetary ephemerides. Moreover, although potentially not free from limitations\footnote{For a recent analysis dealing with a specific effect, see \cite{2012CQGra..29w5027H}.}, the \virg{opportunistic} approach of using existing observation-based determinations, originally obtained for different scopes, to infer bounds on non-standard dynamical effects parameterizing deviations from general relativity has  been widely adopted so far in the literature; see, e.g., the discussion in Section $4.2$ of \cite{2012arXiv1210.3026I} and references therein.

In using the planetary supplementary perihelion precessions $\Delta\dot\varpi$ to gain information on $|\xi K|$, it must be taken into account the fact that, according to \rfr{dOdt1}-\rfr{dodt1} and \rfr{dodt3} in Section \ref{calcolo}, it may occur that, for a given orbital configuration, the predicted anomalous perihelion precessions can vanish for certain particular locations in the sky of $\bds{\hat{K}}$, irrespectively of the actual values of $\xi$ and $K$ for the planet considered. As such, those spatial orientations of $\bds{\hat{K}}$ must be excluded from our analysis since no useful limits on $|\xi K|$ can be inferred. As it turns out from Figure \ref{curve}, the critical positions of $\bds{\hat{K}}$  lie on continuous curves in the $\{\alpha_{\bds{\hat{K}}}, \delta_{\bds{\hat{K}}}\}$ plane, where $\alpha_{\bds{\hat{K}}}$ and $\delta_{\bds{\hat{K}}}$ are the Celestial coordinates of $\bds{\hat{K}}$, i.e. the right ascension (RA) and the declination (DEC), respectively.
\begin{figure*}
\centering
\begin{tabular}{cc}
\epsfig{file=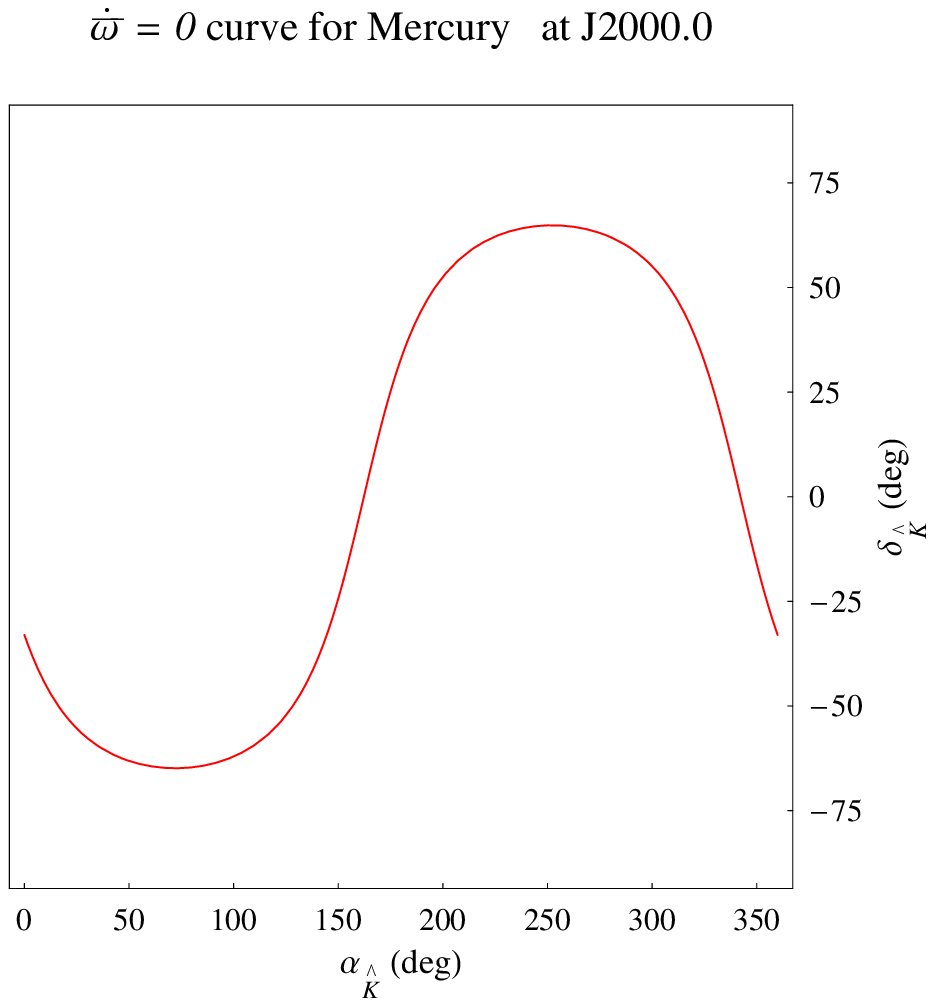,width=0.40\linewidth,clip=} & \epsfig{file=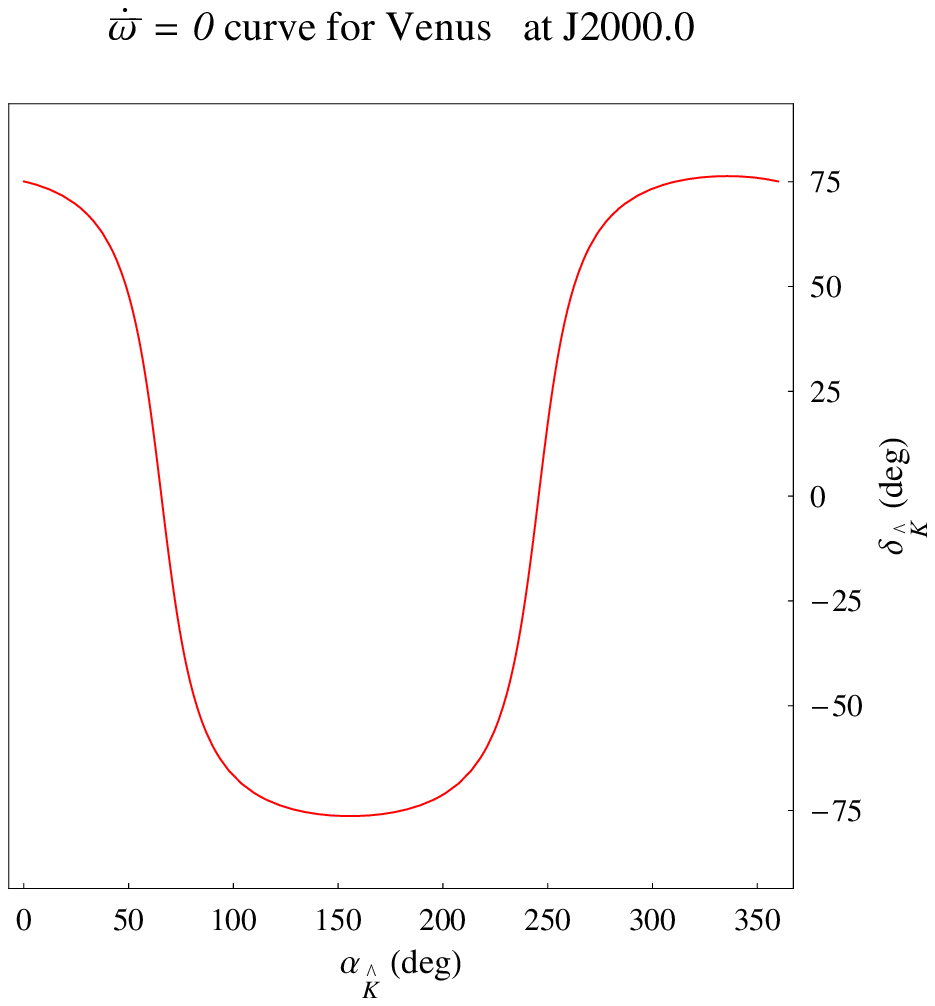,width=0.40\linewidth,clip=}\\
\epsfig{file=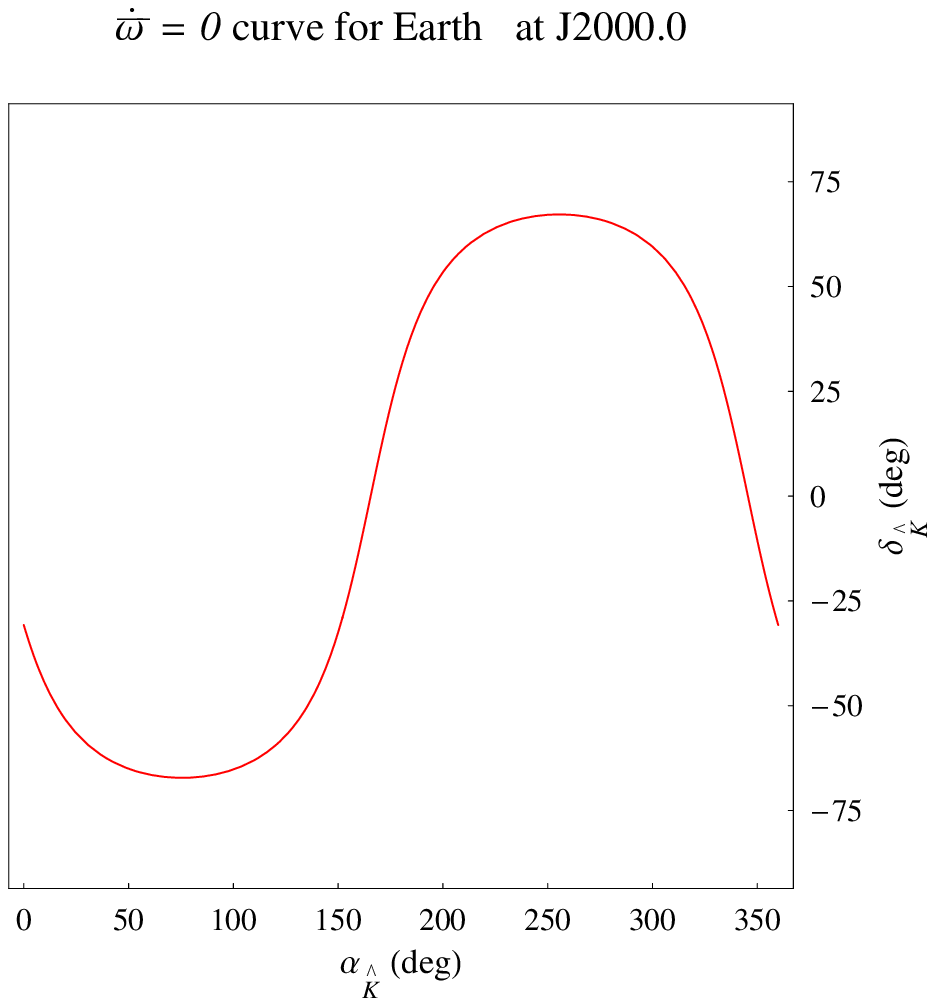,width=0.40\linewidth,clip=} & \epsfig{file=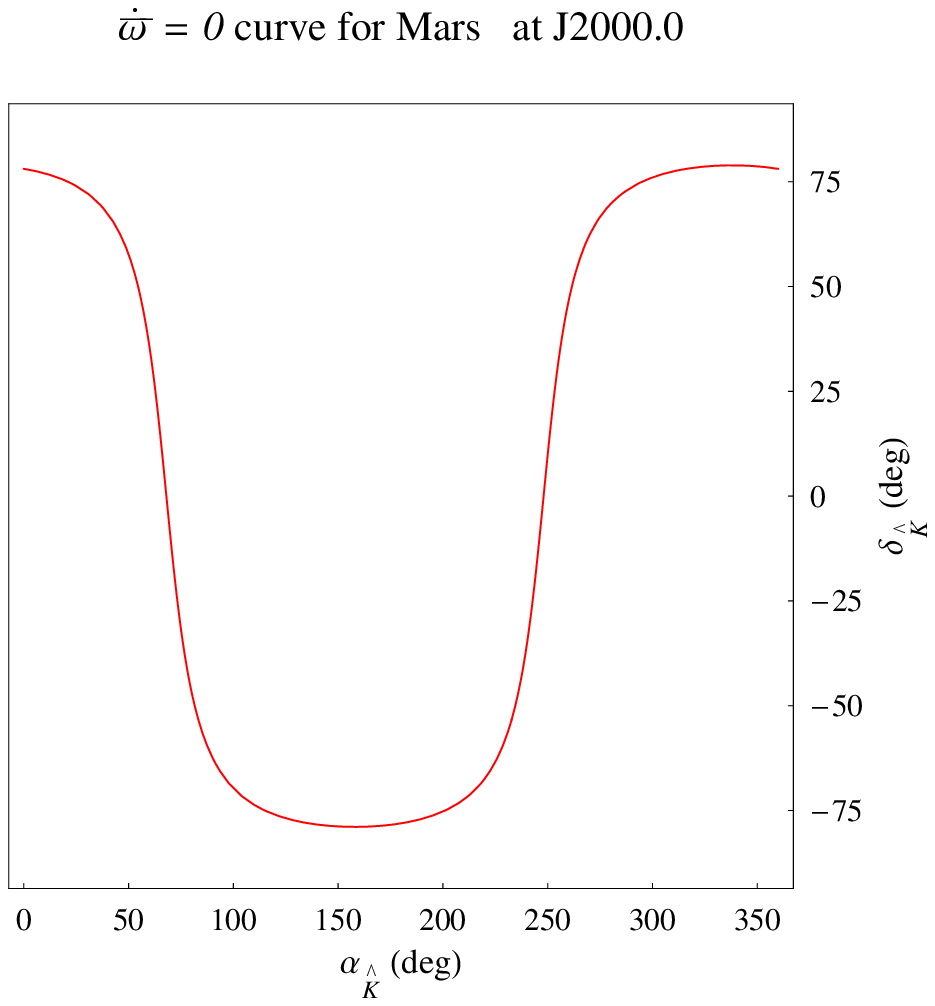,width=0.40\linewidth,clip=}\\
\epsfig{file=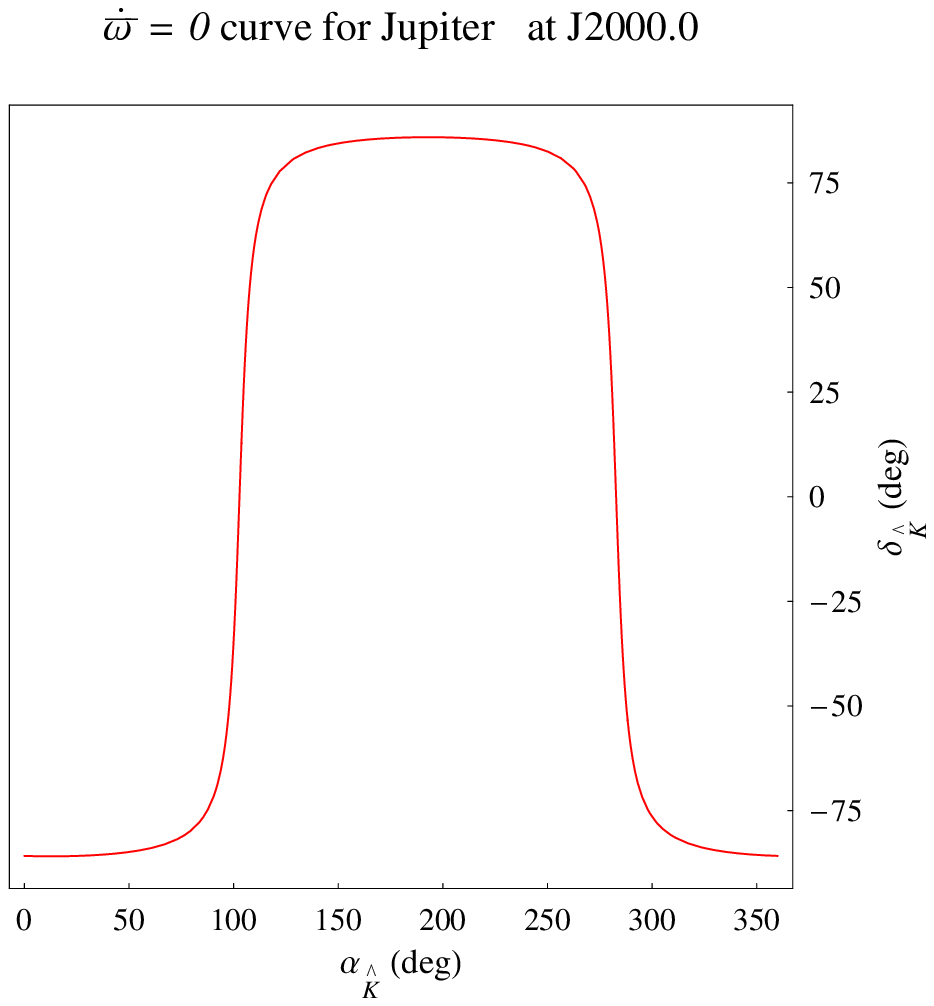,width=0.40\linewidth,clip=} & \epsfig{file=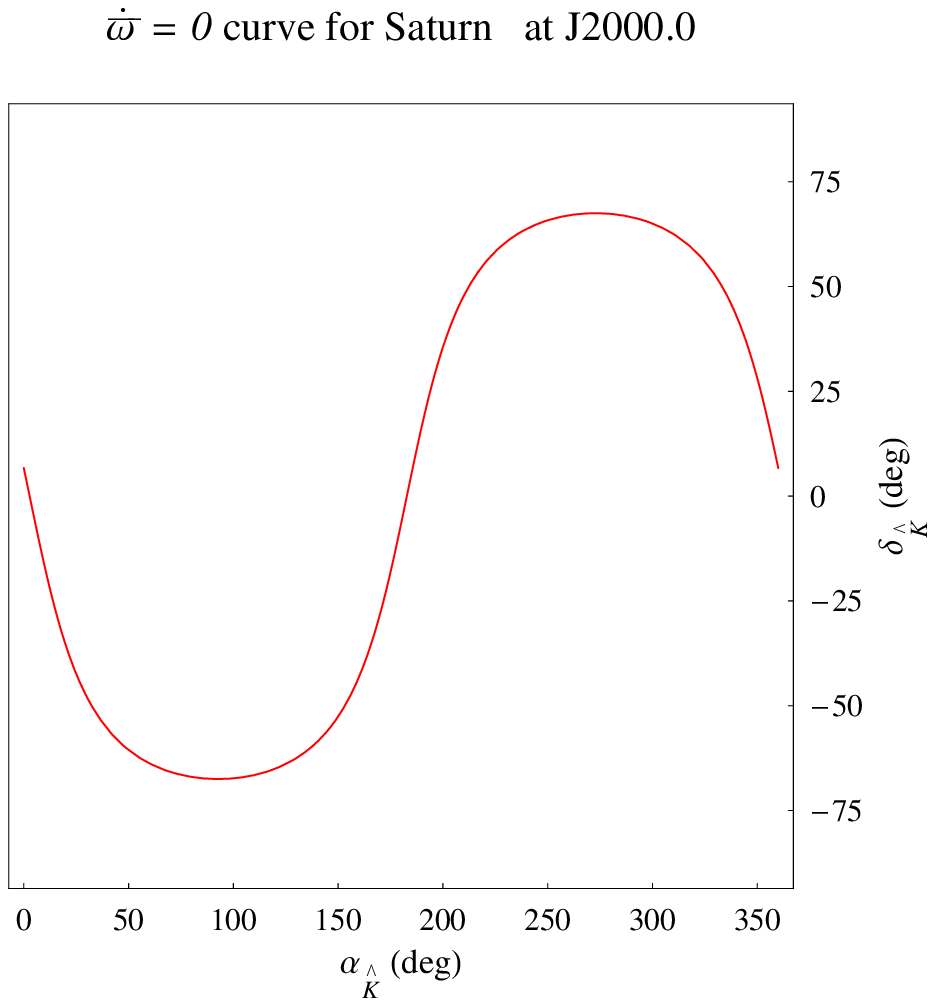,width=0.40\linewidth,clip=}\\
\end{tabular}
\caption{Locations in the sky of $\bds{\hat{K}}$ for which the anomalous perihelion precessions of the first six planets of the solar system, theoretically computed with the formulas of Section \ref{calcolo} and the values of $I,\Omega,\omega$ at the epoch J$2000.0$, vanish.  They form continuous curves in the $\{\alpha_{\bds{\hat{K}}},\delta_{\bds{\hat{K}}}\}$ plane. No useful bounds on $|\xi K|$ can be obtained on them.}\lb{curve}
\end{figure*}
Table \ref{tavola} summarizes the bounds on $|\xi K|$ for all the planets for which the corrections $\Delta\dot\varpi$ are currently available. The smallest value occurs for Mars.
\begin{table*}
\caption{Preliminary bounds on $|K \xi|$ and corresponding locations in the sky of $\bds{\hat{K}}$ inferred from a comparison of the theoretical predictions of Section \ref{calcolo} for the precession of the longitude of pericenter $\varpi$  with the supplementary perihelion rates $\Delta\dot\varpi$ of some planets of the Solar System estimated in \cite{2013MNRAS.432.3431P} with the EPM2011 ephemerides. The values of $I,\Om,\omega$ at the epoch J$2000.0$ were adopted in the formulas of Section \ref{calcolo}. Only standard dynamics was explicitly modeled in the global solution yielding $\Delta\dot\varpi$ \cite{2013MNRAS.432.3431P}. Thus, our results should be regarded just as indicative of what could be obtained with the present state-of-the-art in the field of planetary ephemerides generation. Dedicated covariance analyses by explicitly modeling the non-standard effects treated in this paper may represent a valuable complementary approach to the  one offered here.
}\label{tavola}
\centering
\bigskip
\begin{tabular}{llll}
\hline\noalign{\smallskip}
Planet & $|K\xi|$ (kg s$^{-1}$) & $\alpha_{\bds{\hat{K}}}$ (deg) & $\delta_{\bds{\hat{K}}}$ (deg) \\
\noalign{\smallskip}\hline\noalign{\smallskip}
Mercury & $ 23.5 $ &  $ 72.3 $ & $25.1 $ \\
Venus & $5.0 $ &  $ 155.7 $ & $ 13.6 $ \\
Earth & $ 1.5 $ &  $ 75.5 $ & $ 22.8 $ \\
Mars & $ 0.1 $ &  $ 158 $ & $ 11$ \\
Jupiter & $ 84851 $ &  $12.8 $ & $ 4.1 $ \\
Saturn & $ 368.4 $ &  $92.8 $ & $22.5 $ \\
\noalign{\smallskip}\hline\noalign{\smallskip}
\end{tabular}
\end{table*}

Although the long-term rate of change of the semimajor axis is, perhaps, the most striking consequence of \rfr{accel},   no specifically dedicated studies on planetary long-term evolutions of $a$ and $e$ are currently available in the modern literature. That is, the astronomers \cite{2011CeMDA.111..363F, Pitjevi013, 2013MNRAS.432.3431P} who estimated the corrections $\Delta\dot\varpi$ to the standard perihelion precessions did not explicitly estimate analogous quantities $\Delta\dot a,\Delta\dot e$ for the rates of change of $a$ and $e$. As far as the secular increase of the astronomical unit (au) is concerned \cite{2004CeMDA..90..267K,2005tvnv.conf..163S,2009IAU...261.0702A},  it is now a defining constant whose fixed value is assumed equal to 149 597 870 700 m exactly, as recently decided by the \textit{International Astronomical Union} at its \textit{XXVIII General Assembly} \cite{AU2012}.
Some upper bounds on the planetary secular rates of change of $a$ can only indirectly be inferred by suitably interpreting existing data, as done in Table 1 of \cite{2012CQGra..29q5007I}, based on Table 3 of \cite{2008IAUS..248...20P}. We will use them to get insights on $|\xi K^0|$.
From Section \ref{calcolo}, it turns out that the predicted anomalous rates of $a$ increase with the distance from the Sun, being proportional to $a$ (\rfr{dadt2}) and to $\sqrt{a}$ (\rfr{dadt3}). Since the most accurate bounds on $\dot a_{\rm obs}$ occurred for the rocky planets at the time of the analysis in \cite{2008IAUS..248...20P}, let us consider Mars. A maximum Arean rate of \eqi\dot a_{\rm obs}\lesssim 3\ {\rm m\ cty}^{-1}\eqf is given in Table 1 of \cite{2012CQGra..29q5007I}. Let us, first, evaluate the order-of-magnitude of the contribution of \rfr{dadt3}, which is proportional to $\textcolor{black}{\nk a^2 e \propto} \sqrt{a} e$, to the first order in $e$. It turns out \eqi\rp{\dot a_{\rm nmc}^{\ton{v^2}}}{\xi K}\approx 5\times 10^{-18}\ {\rm m\ kg}^{-1}.\eqf By using the figure quoted in Table \ref{tavola} for $|\xi K|$ of Mars, obtained independently from the Arean perihelion precession, we can conclude that the contribution of \rfr{dadt3} to the expected anomalous rate $\dot a_{\rm nmc}$ is completely negligible for Mars, given the present-day level of accuracy in determining its orbit over centennial timescales. Indeed, it would amount to
\eqi \left|\dot a_{\rm nmc}^{\ton{v^2}}\right| = 5\times 10^{-19}\ {\rm m\ s}^{-1} = 1.6\times 10^{-9}\ {\rm m\ cty}^{-1}.\eqf Thus, we can safely focus just on \rfr{dadt2}, which contains $\xi K^0$ and vanishes if and only if $\xi K^0 = 0$. A comparison of $\dot a_{\rm nmc}^{\ton{v}}$ with the observation-based bound $\dot a_{\rm obs}$ in Table 1 of \cite{2012CQGra..29q5007I} yields
\eqi |\xi K^0|\leq 1249\ {\rm kg\ s^{-1}}. \eqf

\textcolor{black}{
\section{Bounds from Earth's satellites}\lb{terras}
}
\textcolor{black}{
A similar analysis can be performed in the gravitational field of the Earth with the
COBE and GP-B satellites\footnote{See on the Internet \url{http://www.astronautix.com/craft/cobe.htm\#chrono} and \url{http://einstein.stanford.edu/content/fact_sheet/GPB_FactSheet-0405.pdf.}} whose relevant physical and orbital parameters are listed in Table \ref{tavolasats}.
}
\begin{table*}
\caption{Relevant physical and orbital parameters of the COBE  and GP-B satellites. The observation-based bounds on the semimajor axis decays are reported from \protect\cite{2013IJMPA..2850074A} citing private communications by E.N. Wright to the author. The figure for GP-B refers to its drag-free epoch.}\label{tavolasats}
\centering
\bigskip
\begin{tabular}{llllll}
\hline\noalign{\smallskip}
Spacecraft & mass (kg) & $a$ (km) & $e$ & $I$ (deg) & $\dot a_{\rm obs}$ (m s$^{-1}$) \\
\noalign{\smallskip}\hline\noalign{\smallskip}
COBE & $2270$ & $7278$ & $0.00089$ & $98.9$ & $-2\times 10^{-5}$ \\
GP-B & $3100$ & $7027.4$ & $0.0014$ & $90.007$ & $-1\times 10^{-6}$ \\

\noalign{\smallskip}\hline\noalign{\smallskip}
\end{tabular}
\end{table*}
\textcolor{black}{As far as  $|\xi K|$ of the Earth is concerned, an order-of-magnitude hint for it can be preliminarily inferred from the meter-level Root-Mean-Square (RMS) accuracy \cite{2007AdSpR..40....1L} in the orbit determination of GP-B during its yearly drag-free science phase. From \cite{1988JGR....93.6216C}
\eqi\Delta N \approx a\sin I\Delta\Om\eqf for the cross-track shift $\Delta N$, assumed of the order of 5 m, by multiplying \rfr{dOdt1} by 1 yr to have the predicted node shift $\Delta\Om_{\rm nmc}$ allows to obtain
\eqi|\xi K|\approx 2\times 10^{-4}\ {\rm kg\ s^{-1}}.\lb{Terra}\eqf
Although preliminary, the bound of \rfr{Terra} allows to safely use just \rfr{dadt2} for $|\xi K\textcolor{black}{^0}|$. Indeed, it turns out that
\rfr{dadt3}, calculated for GP-B with \rfr{Terra}, yields
\eqi \dot a_{\rm nmc}^{\ton{v^2}}\approx 10^{-8}\ {\rm m\ s^{-1}}, \eqf
which is two orders of magnitude smaller that the observational constraint on the semimajor axis rate quoted in Table \ref{tavolasats}.
As a result, by comparing \rfr{dadt2} with $\dot a_{\rm obs}$ in Table \ref{tavolasats} one gets
\eqi |\xi K\textcolor{black}{^0}|\lesssim 2\times 10^{-10}\ {\rm kg\ s^{-1}}. \lb{wow}\eqf It turns out that COBE yields a bound weaker than \rfr{wow} by one order of magnitude.
It must be noted how the \rfr{Terra} and \rfr{wow} for the Earth are tighter than those obtained for the Sun in Section \ref{soles} by several orders of magnitude.
}
\section{Summary and overview}
In this paper, we looked at the effects of a non-geodesic, \virg{pressure}-type acceleration on the orbital motion of a \textcolor{black}{structureless massive} test particle in a localized, gravitationally bound two-body system. It arises from a general class of modified gravitational theories with a nonminimal coupling  depending generally on the curvature of the spacetime.
In particular, we considered a recently published large class of theories in which the coupling can be  a quite general function of the set of 9 parity-even curvature invariants.

 It turned out that all the usual Keplerian orbital elements  experience long-term variations. In our calculation, we did not make any a-priori simplifying assumptions on both the orbital configuration of the test particle and on the spatial orientation of $\bds K$. Nonetheless, we assumed that $\mathfrak{m},\xi,K^{\alpha}$ can be considered constant over one orbital period of the test particle. It should be fully justified in future, if and when the theory considered  will be studied in more details for the specific case of a two-body system. \textcolor{black}{In particular, theorists should at least attempt to solve the spherically symmetric metric generated by an isolated mass to check if the terms with  fewest number of derivatives  in the nonminimal couplings become dominant, thus introducing new mass scales in the problem.}. Among the other precessions, both the semimajor axis $a$ and the eccentricity $e$, which characterize the mean distance in a two-body system, do not stay generally constant.  It was also confirmed by a numerical integration of the equations of motion of a fictitious Sun-planet system. Interestingly, such a peculiar feature yields a mechanism which, in principle, has the potential capability of explaining certain puzzles concerning the history of the ancient Earth, like the Faint Young Sun Paradox, in terms  of a steady orbital recession of our planet away from the Sun.

 We used our analytical predictions to tentatively pose some preliminary bounds on $\xi K^0$ and $\xi K$, treated as free parameters to be potentially constrained from observations, by using the latest determinations in the field of the planetary ephemerides of the solar system. The perihelion $\varpi$ and the semimajor axis $a$ of Mars preliminarily yielded  $|\xi K|\lesssim 0.1$ kg s$^{-1}$ and $|\xi K^0|\lesssim 1249$ kg s$^{-1}$, respectively. Strictly speaking, our results may not be considered as genuine constraints. Indeed, they did not come from a least-square fitting procedure of dynamical models including the effects treated here to the existing planetary data record. In fact, the observation-based determinations used by us were estimated by the astronomers by only including  standard dynamics. Thus, our bounds should be viewed as indications of acceptable values from the most recent ephemerides.
\textcolor{black}{
An analogous analysis for the Earth's scenario performed with the artificial satellites COBE and GP-B yields much tighter bounds. Indeed, we have $|\xi K|\lesssim 2\times 10^{-4}$ kg s$^{-1}$ and $|\xi K\textcolor{black}{^0}|\lesssim 2\times 10^{-10}$ kg s$^{-1}$, respectively.
}
\section*{Acknowledgments}
I thank D. Puetzfeld for his helpful clarifications on the covariance of \rfr{4accel} and on the nature of $\mathfrak{m},\xi,K^{\alpha}$. I am also grateful to three anonymous referees for their insightful comments.

\bibliography{Nonminimalbib}{}

\end{document}